\documentclass[preprint,aps,nofootinbib,preprintnumbers,amsmath,amssymb]{revtex4-1}
\usepackage{epsfig}
\usepackage{subfigure}
\usepackage{dcolumn}
\usepackage{bm}
\usepackage{slashed}
\usepackage{float}
\usepackage{subfigure}
\usepackage{amsmath}
\usepackage{slashed}
\usepackage{nonfloat}
\usepackage{multirow}
\begin{document}
\title{  Semileptonic weak decays of anti-triplet charmed baryons in the light-front formalism}

\author{C.Q. Geng$^{1,2,3}$, Chia-Wei Liu$^{1,2,3}$ and Tien-Hsueh Tsai$^{3}$}
\affiliation{
$^{1}$School of Fundamental Physics and Mathematical Sciences, Hangzhou Institute for Advanced Study, UCAS, Hangzhou 310024, China \\
$^{2}$International Centre for Theoretical Physics Asia-Pacific, Beijing/Hangzhou, China \\
$^{3}$Department of Physics, National Tsing Hua University, Hsinchu 300, Taiwan
}\date{\today}

\begin{abstract}
We systematically study the semileptonic decays of ${\bf B_c} \to {\bf B_n}\ell^+ \nu_{\ell}$ in the light-front constituent quark model, where ${\bf B_c}$ represent the anti-triplet charmed baryons of $(\Xi_c^0,\Xi_c^+,\Lambda_c^+)$ and ${\bf B_n}$ correspond to the octet ones.  We determine the spin-flavor structures of  the constituents in the baryons  with the Fermi statistics  and calculate the decay branching ratios (${\cal B}$s)  and averaged asymmetry parameters ($\alpha$s) with the helicity formalism.  In particular, we find that 
 ${\cal B}( \Lambda_c^+ \to \Lambda e^+ \nu_{e}, ne^+ \nu_{e})=(3.55\pm1.04, 0.36\pm0.15)\%$, 
 ${\cal B}( \Xi_c^+ \to \Xi^0 e^+ \nu_{e},\Sigma^0 e^+ \nu_{e},\Lambda e^+ \nu_{e})=(11.3\pm3.35), 0.33\pm0.09,0.12\pm0.04\%$ 
 and  ${\cal B}( \Xi_c^0 \to \Xi^- e^+ \nu_{e},\Sigma^- e^+ \nu_{e})=(3.49\pm0.95,0.22\pm0.06)\%$. 
 Our results  agree with the current experimental data. Our prediction for  ${\cal B}( \Lambda_c^+ \to  n e^+ \nu_{e})$ is consistent with those in the literature, 
which can be measured by the  charm-facilities, such as BESIII and BELLE. Some of our results for  the  $\Xi_c^{+(0)}$ semileptonic channels can be tested by the experiments at  BELLE as well as the ongoing ones at LHCb and  BELLEII. 

\end{abstract}
\maketitle

\section{introduction}
In the recent few  years, there have been two important experiments in charmed baryon physics. 
One is the  measurement of the absolute branching ratios of ${\cal B} (\Xi_c^0 \to \Xi^- \pi^+)=(1.8\pm0.5 )\%$~\cite{Li:2018qak} and
${\cal B} (\Xi_c^+ \to \Xi^- \pi^+\pi^+)=(2.86 \pm 1.21 \pm 0.38)\%$~\cite{Li:2019atu} from the Belle Collaboration, and the other is the most precise measurement of the $\Xi_c^0$'s lifetime of $\tau_{\Xi_c^0}=154.5 \pm 1.7 \pm 1.6 \pm 1.0 \text{ fs}$ from the LHCb Collaboration~\cite{Aaij:2019lwg}, which significantly deviates from the past world averaged value of $\tau_{\Xi_c^0}=112^{+13}_{-10}~\text{fs}$ in PDG~\cite{Zyla:2020zbs}. Both of them bring us new hints as well as new problems  in charm physics. We are now in a precision-era of charm physics. It is  expected that as more high quality data will be accumulated in the future,  
stronger constraints on various baryonic QCD models  as well as  physics beyond the standard model can be given.
 
Recently,  the anti-triplet charm baryon decays
 have been extensively discussed in the literature.
 However,  due to the large non-perturbative effects from the quantum chromodynamics (QCD),  the decay amplitudes and observables of hadrons
are very hard to obtain from the QCD first principle. 
To avoid the difficulties, various charmed baryon decay processes 
have been studied based on the  flavor symmetry of $SU(3)_f$, such as  semi-leptonic, two-body and three-body non-leptonic decays, to get reliable results~\cite{Geng:2017mxn,Savage:1989qr,Savage:1991wu,zero,He:2018joe,He:2018php,Wang:2017azm,Geng:2018plk,Geng:2018rse,Geng:2018upx,Geng:2019awr,Geng:2019bfz,Grossman:2019xcj,Lu:2016ogy,Wang:2017gxe,Cen:2019ims,Hsiao:2019yur,Roy:2019cky,Jia:2019zxi,Geng:2020fng}.
It is known that 
  $SU(3)_f$ is an approximate symmetry, resulting in about $10\%$ of uncertainties for the predictions  inherently. 
  Moreover, the $SU(3)_f$ symmetry itself does not directly reveal any clue about the QCD dynamics.
  In fact, its applications  heavily rely  on the experimental data as inputs. In order to have precise calculations,
 we need a specific dynamical QCD model to understand each decay process.
%
For simplicity, we only discuss the semi-leptonic processes of the anti-triplet charmed baryons in this study, which involve  purely  factorizable contributions.
There are several theoretical calculations on these  decay processes  with different QCD frameworks   
in the literature~\cite{Cheng:1995fe,Zhao:2018zcb,PerezMarcial:1989yh,Faustov:2019ddj,Faustov:2016yza,Gutsche:2015rrt,Meinel:2016dqj,Meinel:2017ggx,Azizi:2009wn,Liu:2009sn}. 

The light front (LF) QCD formalism in the quark model is a consistent relativistic approach, which has been tested  successfully  
in the mesonic and light quark sectors in early times~\cite{Schlumpf:1992ce,Zhang:1994ti}. 
Because of these successes, it has been used in other generalized systems, such as those containing the heavy mesons, pentaquarks and so on~\cite{Cheng:2004cc,Chang:2019obq,Zhao:2018mrg,Xing:2018lre,Cheng:2017pcq,Geng:2000fs,Geng:2001vy,
Geng:2016pyr}.
Apart from the charm system, the bottom  to charmed baryon nonleptonic decays have been recently
analyzed in the LF formalism~\cite{Chua:2018lfa}.
For a review on the comprehensive introduction of the LF QCD and its vacuum structure, one can refer to  Ref.~\cite{Zhang:1994ti}.
 For  the LF constituent quark model (LFCQM), we recommend  Ref.~\cite{Schlumpf:1992ce} and references therein.
The  advantage of LFCQM  is that we
can boost the reference frame without changing the equation of motion because of the commutativity of the LF Hamiltonian and  boost generators. 
 It provides us with a great convenience to calculate the wave-function in different inertial frames because of the recoil effects in the  transition form factors.

This paper is organized as follows.  We first present our formal calculations of the branching ratios and averaged decay asymmetries in terms of 
the helicity amplitudes, the baryonic states  in LFCQM, and the baryonic transition form factors in Sec.~II.
  In Sec. III, we show our numerical results  and compare them with those in the literature.
  In Sec.~IV, we give our   conclusions.

\section{Formalism}

\subsection{Helicity Amplitudes and Observables }
The effective Hamiltonian for the anti-triplet charmed baryon semi-leptonic weak decays can be written as
\begin{eqnarray}
	{\cal H}_{eff}=\frac{G_F}{\sqrt{2}}V_{cq}(\bar{\nu_{\ell}}\ell)_{V-A}(\bar{q}c)_{V-A}
\end{eqnarray}
where $G_F$ is the Fermi constant, $V_{cq}$ is the CKM matrix element,  and
$(\bar{\ell}\nu_{\ell})_{V-A}$ and $(\bar{q}c)_{V-A}$   denote as the usual $V-A$ currents $\bar{\ell}\gamma^\mu (1-\gamma_5)\nu_{\ell}$  and $\bar{q}\gamma_\mu (1-\gamma_5)c$ with $q=d,s$, respectively.
The weak transition amplitudes of the  anti-triplet charmed baryons are given by
\begin{eqnarray}
{\cal A}({\bf B}_c\to {\bf B}_n \ell^+\nu_{\ell})=\frac{G_F}{\sqrt{2}}V_{cq}\bar{\ell}\gamma^\mu (1-\gamma_5)\nu_{\ell}\langle {\bf B}_n |\bar{q}\gamma_\mu (1-\gamma_5)c|{\bf B}_c \rangle
\end{eqnarray}
where
 the baryon transition matrix elements are parameterized by 
\begin{eqnarray}
&&\langle {\bf B}_n ,p_f,S_z'|\bar{q}\gamma_\mu (1-\gamma_5)c|{\bf B}_c ,p_i,S_z\rangle \nonumber\\
&&=\bar{u}_{{\bf B}_n}(p_f,S_z')\left[ \gamma_\mu f_1(k^2)-i\sigma_{\mu \nu}\frac{k^{\nu}}{M_{{\bf B}_c}}f_2(k^2)+f_3(k^2)\frac{k_{\mu}}{M_{{\bf B}_c}}\right]u_{{\bf B}_c}(p_i,S_z)\nonumber\\
&&-\bar{u}_{{\bf B}_n}(p_f,S_z')\left[ \gamma_\mu g_1(k^2)-i\sigma_{\mu \nu}\frac{k^{\nu}}{M_{{\bf B}_c}}g_2(k^2)+g_3(k^2)\frac{k_{\mu}}{M_{{\bf B}_c}}\right]\gamma_5u_{{\bf B}_c}(p_i,S_z)
\end{eqnarray}
with $k^{\mu}=p_i^{\mu}-p_f^{\mu}$,  $\sigma_{\mu \nu}=\frac{i}{2}\left[\gamma_\mu,\gamma_\nu\right]$, and $f_i(k^2)$ and $g_i(k^2)$ being
the form factors describing the  non-perturbative QCD effect in the time-like range of $m_{\ell}^2<k^2<(M_{\bf B_c}-M_{\bf B_n})^2$.
We introduce a set of  helicity amplitudes $H^{V(A)}_{\lambda_2\lambda_W}$ to calculate the decay branching ratios and other physical quantities,
 where $\lambda_2$ and $\lambda_W$ represent the helicity quantum numbers of the daughter baryon and  off-shell $W^+$ boson in the decay processes,
  respectively.  These amplitudes  give more intuitive physical pictures about the helicity structures of  the  decay processes. Furthermore, when we 
  evaluate the asymmetries of these processes, such as  the integrated (averaged) decay asymmetry, also known as the longitudinal polarization of the daughter baryon, these amplitudes result in much simpler expressions than the traditional ones. 
Relations between the helicity amplitudes and form factors are given by\cite{Kadeer:2005aq,Zhao:2018zcb}
\begin{eqnarray}
H^{V}_{\frac{1}{2}1}&=&\sqrt{2K_-}\left(-f_1(k^2)-\frac{M_{\bf B_{c}}+M_{\bf B_{n}}}{M_{\bf B_{c}}}f_2(k^2) \right) \,,\nonumber \\
H^{V}_{\frac{1}{2}0}&=&\frac{\sqrt{K_-}}{\sqrt{k^2}}\left((M_{\bf B_{c}}+M_{\bf B_{n}})f_1(k^2)+\frac{k^2}{M_{\bf B_{c}}}f_2(k^2)\right)\,,\nonumber \\
H^{V}_{\frac{1}{2}t}&=&\frac{\sqrt{K_+}}{\sqrt{k^2}}\left((M_{\bf B_{c}}+M_{\bf B_{n}})f_1(k^2)+\frac{k^2}{M_{\bf B_{c}}}f_3(k^2)\right)\,,\nonumber \\
H^{A}_{\frac{1}{2}1}&=&\sqrt{2K_+}\left(-g_1(k^2)-\frac{M_{\bf B_{c}}+M_{\bf B_{n}}}{M_{\bf B_{c}}}g_2(k^2)\right)\,,\nonumber \\
H^{A}_{\frac{1}{2}0}&=&\frac{\sqrt{K_+}}{\sqrt{k^2}}\left((M_{\bf B_{c}}-M_{\bf B_{n}})g_1(k^2)-\frac{k^2}{M_{\bf B_{c}}}g_2(k^2)\right) \,,\nonumber\\
H^{A}_{\frac{1}{2}t}&=&\frac{\sqrt{K_-}}{\sqrt{k^2}}\left((M_{\bf B_{c}}-M_{\bf B_{n}})g_1(k^2)-\frac{k^2}{M_{\bf B_{c}}}g_3(k^2)\right)\,,
\label{amp}
\end{eqnarray} 
where $K_{\pm}=(M_{\bf B_c}\pm M_{\bf B_n})^2-k^2$. 

The differential decay widths and asymmetries can be expressed in the following analytic forms in terms of the helicity amplitudes with 
the non-vanishing lepton masses of  $m_{\ell}$~\cite{Geng:2019bfz},
\begin{eqnarray}\label{Br}
	\frac{d \Gamma}{d k^{2}}&= &\frac{1}{3} \frac{G_{F}^{2}}{(2 \pi)^{3}}\left|V_{q c}\right|^{2} \frac{\left(k^{2}-m_{\ell}^{2}\right)^{2} p}{8 M_{\bf{B}_{c}}^{2} k^{2}}\left[\left(1+\frac{m_{\ell}^{2}}{2 k^{2}}\right)\right. \nonumber\\
	&&\left.\left(\left|H_{\frac{1}{2} 1}\right|^{2}+\left|H_{-\frac{1}{2}-1}\right|^{2}+\left|H_{\frac{1}{2}0}\right|^{2}+\left|H_{-\frac{1}{2} 0}\right|^{2}\right)+\frac{3 m_{\ell}^{2}}{2 k^{2}}\left(\left|H_{\frac{1}{2} t}\right|^{2}+\left|H_{-\frac{1}{2} t}\right|^{2}\right)\right]\,,
\end{eqnarray}
\begin{scriptsize}
\begin{eqnarray}\label{asy}
\langle\alpha( k^{2})\rangle= \frac{\int dk^2 \frac{\left(k^{2}-m_{\ell}^{2}\right)^{2} p}{8 M_{\bf{B}_{c}}^{2} k^{2}}\left[\left(1+\frac{m_{\ell}^{2}}{2 k^{2}}\right)\right. 
\left.\left(\left|H_{\frac{1}{2} 1}\right|^{2}-\left|H_{-\frac{1}{2}-1}\right|^{2}+\left|H_{\frac{1}{2}0}\right|^{2}-\left|H_{-\frac{1}{2} 0}\right|^{2}\right)+\frac{3 m_{\ell}^{2}}{2 k^{2}}\left(\left|H_{\frac{1}{2} t}\right|^{2}-\left|H_{-\frac{1}{2} t}\right|^{2}\right)\right]}{\int dk^2\frac{\left(k^{2}-m_{\ell}^{2}\right)^{2} p}{8 M_{\bf{B}_{c}}^{2} k^{2}}\left[\left(1+\frac{m_{\ell}^{2}}{2 k^{2}}\right)\right. 
\left.\left(\left|H_{\frac{1}{2} 1}\right|^{2}+\left|H_{-\frac{1}{2}-1}\right|^{2}+\left|H_{\frac{1}{2}0}\right|^{2}+\left|H_{-\frac{1}{2} 0}\right|^{2}\right)+\frac{3 m_{\ell}^{2}}{2 k^{2}}\left(\left|H_{\frac{1}{2} t}\right|^{2}+\left|H_{-\frac{1}{2} t}\right|^{2}\right)\right]}\,,
\end{eqnarray}
\end{scriptsize}
where $p=\sqrt{K_+K_-}/{2M_{\bf B _c}}$ and $H_{\lambda_2 \lambda_W}=H_{\lambda_2 \lambda_W}^{V}-H_{\lambda_2 \lambda_W}^{A}$.

\subsection{Light Front Constituent Quark Model}
From Eqs.~(\ref{Br}) and (\ref{asy}), we see that as long as  the time-like form factors are known, both branching ratios and averaged decay asymmetries can be determined. To calculate these time-like form factors, we use  LFCQM,  in which 
 a baryon is treated as a bound state of three constituent quarks quantized in the LF formalism and its state is denoted by the momentum $P$, canonical spin $S$ and the z-direction projection of spin $S_z$, respectively. 
As a result, the baryon state can be expressed by~\cite{Zhang:1994ti,Schlumpf:1992ce,Cheng:2004cc,Ke:2007tg,Ke:2012wa,Ke:2019smy}
\begin{eqnarray}
\label{eq1}
	&|{\bf B}&,P,S,S_z\rangle=\int\{{d^3{\tilde{p}}}\}2(2\pi)^3\frac{1}{\sqrt{P^+}}\delta^3(\tilde{P}-\tilde{p}_1-\tilde{p}_2-\tilde{p}_3) \nonumber \\
	&\times& \sum_{\lambda_1,\lambda_2,\lambda_3}\Psi^{SS_z}(\tilde{p}_1,\tilde{p}_2,\tilde{p}_3,\lambda_1,\lambda_2,\lambda_3)C^{\alpha\beta\gamma}F_{abc}|q_{\alpha}^{a}(\tilde{p}_1,\lambda_1) q_{\beta}^{b}(\tilde{p}_2,\lambda_2)q_{\gamma}^{c}(\tilde{p}_3,\lambda_3) \rangle\,,
	\label{baryon}
\end{eqnarray}
where  $\Psi^{SS_z}(\tilde{p}_1,\tilde{p}_2,\tilde{p}_3,\lambda_1,\lambda_2,\lambda_3)$  is
 the vertex function describing the overlapping between the baryon and  its constituents, which can be formally solved from the three-body Bethe-Salpeter equations,
$C^{\alpha\beta\gamma}$ ($F_{abc}$) are the color (flavor) factors, $\lambda_i$  and $\tilde{p}_i$ with $i=1,2,3$  are the LF helicities and
3-momenta of the on-mass-shell constituent quarks, defined as
\begin{eqnarray}
\tilde{p}_{i}=(p_i^+,p_{i\perp})\,, ~p_{i\perp}=(p_i^1,p_i^2) \,,~ p_i^-=\frac{m_i^2+p_{i\perp}^2}{p_i^+} \,,
\end{eqnarray}
and $|q_\alpha^a(\tilde{p},\lambda)\rangle$ and $\{{d^3\tilde{p}}\}$ correspond to
the light front constituent quark states and the integral measure, given by
\begin{eqnarray}
	 |q_\alpha^a(\tilde{p},\lambda)\rangle=d^{\dagger a}_{\alpha}(\tilde{p},\lambda)|0\rangle\,,~\{{d^3\tilde{p}}\}\equiv\prod_{i=1,2,3} \frac{dp_i^+d^2p_{i\perp}}{2(2\pi)^3}\,, 
	\end{eqnarray}
	respectively, with the quark field operators  satisfied the following anti-commutation relations
	\begin{eqnarray}
	 ~\{d_{\alpha'}^{a'}(\tilde{p'},\lambda'),d^{\dagger a}_{\alpha}(\tilde{p},\lambda)\}=2(2\pi)^{3}\delta^3(\tilde{p'}-\tilde{p})\delta_{\lambda'\lambda}\delta_{\alpha'\alpha}\delta^{a'a}\,,
	 \delta^3(\tilde{p})=\delta(p^+)\delta^2(p_{\perp})\,.
	\label{state}
\end{eqnarray}
To separate the internal motion of the constituents from the bulk motion, we use the  kinematic variables of $(q_{\perp},\xi)$,
 $(Q_{\perp},\eta)$ and $P_{tot}$, given by
\begin{eqnarray}
P_{tot}&=&\tilde{P}_1+\tilde{P}_2+\tilde{P}_3, \qquad \xi=\frac{p_1^+}{p_1^++p_2^+}, \qquad \eta=\frac{p_1^++p_2^+}{P_{tot}^+}\,,
 \nonumber \\
q_{\perp}&=&(1-\xi)p_{1\perp}-\xi p_{2\perp},\quad  Q_{\perp}=(1-\eta)(p_{1\perp}+p_{2\perp})-\eta p_{3\perp} \,,
\label{Lkin}
\end{eqnarray}
where $(q_{\perp},\xi)$ and $(Q_{\perp},\eta)$ capture the relative motions between the first and second quarks,
 and the third  and other two quarks, respectively. 
We consider the three constituent quarks in the baryon independently with suitable spin-flavor structures satisfying the Fermi statistics
to have a correct baryon bound state system. 
The vertex function of $\Psi^{SS_z}(\tilde{p}_1,\tilde{p}_2,\tilde{p}_3,\lambda_1,\lambda_2,\lambda_3)$ in Eq.~(\ref{eq1}) 
can  be further written into two parts~\cite{Lorce:2011dv,Zhang:1994ti,Schlumpf:1992ce},
\begin{eqnarray}
\Psi^{SS_z}(\tilde{p}_1,\tilde{p}_2,\tilde{p}_3,\lambda_1,\lambda_2,\lambda_3)&=&\phi(q_\perp,\xi,Q_\perp,\eta)\Xi^{SS_z}(\lambda_1,\lambda_2,\lambda_3)\,,
\end{eqnarray}
where $\phi(q_\perp,\xi,Q_\perp,\eta)$ is
the momentum distribution of constituent quarks  and $\Xi^{SS_z}(\lambda_1,\lambda_2,\lambda_3)$ represents
the momentum-dependent  spin wave function, given by 
\begin{eqnarray}
\Xi^{SS_z}(\lambda_1,\lambda_2,\lambda_3)&=&\sum_{s_1,s_2,s_3}\langle\lambda_1|R^{\dagger}_1|s_1\rangle\langle\lambda_2|R^{\dagger}_2|s_2\rangle\langle\lambda_3|R^{\dagger}_3|s_3\rangle \bigg\langle\frac{1}{2}s_1,\frac{1}{2}s_2,\frac{1}{2}s_3\bigg|SS_z\bigg\rangle\,,
\end{eqnarray}
with  the $SU(2)$ Clebsch-Gordan coefficients of  $\big\langle\frac{1}{2}s_1,\frac{1}{2}s_2,\frac{1}{2}s_3\big|SS_z\big\rangle$,  and  $R_i$  is the Melosh transformation matrix~\cite{Polyzou:2012ut,Schlumpf:1992ce},  which corresponds to the $i$th constituent quark, expressed by
\begin{eqnarray}
R_M(x,p_{\perp},m,M)&=&\frac{m+xM-i\vec{\sigma}\cdot(\vec{n}\times \vec{q})}{\sqrt{(m+xM)^2+q_\perp^2}}\,,
\nonumber\\
&&R_1=R_M(\eta,Q_\perp,M_3,M)R_M(\xi,q_\perp,m_1,M_3)\,, \nonumber\\
&&R_2=R_M(\eta,Q_\perp,M_3,M)R_M(1-\xi,-q_\perp,m_2,M_3) \,,\nonumber\\
&&R_3=R_M(1-\eta,-Q_\perp,m_3,M)\,.
\end{eqnarray}
Here, $\vec{\sigma}$ stands for the Pauli matrix,  $\vec{n}=(0,0,1)$, and $M$ and $M_3$ are invariant masses of  $(q_{\perp},\xi)$ and $(Q_{\perp},\eta)$ 
systems, represented by~\cite{Schlumpf:1992ce}
\begin{eqnarray}
M_3^2=\frac{q_\perp^2}{\xi(1-\xi)}+\frac{m_1^2}{\xi}+\frac{m_2^2}{1-\xi}\,, \nonumber\\
M^2=\frac{Q_\perp^2}{\eta(1-\eta)}+\frac{M_3^2}{\eta}+\frac{m_3^2}{1-\eta}\,,
\end{eqnarray}
respectively.

The spin-flavor structures of ${\bf B_c}$ and ${\bf B_n}$ are given by
\begin{eqnarray}
&&|{\bf B_c}\rangle=\frac{1}{\sqrt{6}}[\phi_3\chi^{\rho3}(|q_1q_2c\rangle-|q_2q_1c\rangle)+\phi_2\chi^{\rho2}(|q_1cq_2\rangle-|q_2cq_1\rangle)+\phi_1\chi^{\rho1}(|cq_1q_2\rangle-|cq_2q_1\rangle)]\,,
\nonumber\\
&&|\Lambda^0\rangle=\frac{1}{\sqrt{6}}\phi[\chi^{\rho3}(|dus\rangle-|uds\rangle)+\chi^{\rho2}(|dsu\rangle-|usd\rangle)+\chi^{\rho1}(|sdu\rangle-|sud\rangle)]\,,
\nonumber\\
&&|\Sigma^0\rangle=\frac{1}{\sqrt{6}}\phi[\chi^{\lambda 3}(|dus\rangle+|uds\rangle)+\chi^{\lambda2}(|dsu\rangle+|usd\rangle)+\chi^{\lambda 1}(|sdu\rangle+|sud\rangle)]\,,
\nonumber\\
&&| B_n\rangle=\frac{1}{\sqrt{3}}\phi[\chi^{\lambda_3}|q_1q_1q_2\rangle+\chi^{\lambda_2}|q_1q_2q_1\rangle+\chi^{\lambda_1}|q_2q_1q_1\rangle]\,,
\end{eqnarray}		
where $q_{1(2)}=u,d,s$, $B_n=(N,\Xi,\Sigma^+,\Sigma^-)$, $\phi_{(i)}$ is the momentum distribution  of the constituents with 
the corresponding  spin-flavor  configuration, and
\begin{eqnarray}		
\chi^{\rho 3}_{\uparrow}&=&\frac{1}{\sqrt{2}}(|\uparrow\downarrow\uparrow\rangle-|\downarrow\uparrow\uparrow\rangle)\,,
\quad \chi^{\lambda 3}_{\uparrow}=\frac{1}{\sqrt{6}}(|\uparrow\downarrow\uparrow\rangle+|\downarrow\uparrow\uparrow\rangle-2|\uparrow\uparrow\downarrow\rangle)\,,
\end{eqnarray}
In principle, one could solve $\phi_{(i)}$ from the Bethe-Salpeter equation with an explicit QCD-inspired potential but it is beyond the scope of this paper. Nonetheless, we use a Gaussian type distribution with the phenomenological shape parameters $\beta_{Q}$ and $\beta_{q}$ to describe 
the relative motions of  constituents. Consequently, we  represent the LF kinematic variables $(\xi,q_\perp)$ and $(\eta,Q_\perp)$ in the forms of  ordinary 3-momenta ${\bf q}=(q_{\perp},q_z)$ and  ${\bf Q}=(Q_{\perp},Q_z)$: 
\begin{eqnarray}		
\phi&=&{\cal N}\sqrt{\frac{\partial q_z}{\partial \xi}\frac{\partial Q_z}{\partial \eta}}e^{-\frac{{\bf Q}^2}{2\beta_{Q}^2}-\frac{{\bf q}^2}{2\beta_q^2}}\,, 	 \nonumber\\
q_z&=&\frac{\xi M_3}{2}-\frac{m_1^2+q_\perp^2}{2M_3\xi}\,, \quad Q_z=\frac{\eta M}{2}-\frac{M_3^2+Q_\perp^2}{2M\eta} \,,
\label{3kin}
\end{eqnarray}
where $\cal N$ is the normalization constant.
 Since the one-particle baryonic  state is normalized as
\begin{eqnarray}
\langle {\bf B}&,P',S',S'_z|{\bf B}&,P,S,S_z\rangle=2(2\pi)^3P^+\delta^3(\tilde{P'}-\tilde{P})\delta_{S_z'S_z}\,,
\label{baryonN}
\end{eqnarray}
 the normalization condition of the momentum wave function is given by
\begin{eqnarray}
\frac{1}{2^2(2\pi)^6}	\int d\xi d\eta d^2q_{\perp}d^2Q_{\perp} |\phi_{(3)}|^2=1\,.
\end{eqnarray} 
 In this paper, we take different shape parameters of $\beta_q$ and $\beta_{Q}$  in the momentum wave functions $\phi_i$ to describe the scalar diquark effects in ${\bf B_c}$.  
On the other hand, we assume the momentum-distribution function $\phi$ of octet baryons ${\bf B_n}$ is flavor symmetric for all constituents. 
In other words, the $SU(3)_f$ flavor symmetry is hold in the momentum wave function of ${\bf B_n}$. As a result, the shape parameters of $\phi$ are equal, $i.e.$ $\beta_{Q{\bf B_n}}=\beta_{q{\bf B_n}}=\beta_{{\bf B_n}}$.
 Note that there is no $SU(6)$ spin-flavor symmetry in ${\bf B_c}$ and ${\bf B_n}$  because of the momentum-dependent Melosh transformation even though the forms of these states are similar to those with the  $SU(6)$ spin-flavor wave functions. 
 
\subsection{Transition form factors}
We pick the $\bar{q}\gamma^{+}(1-\gamma_5)c$ current or so-called ``good component'' of the baryon transition amplitudes
\begin{eqnarray}
&&\langle {\bf B}_n ,p_f,S_z'|\bar{q}\gamma^+ (1-\gamma_5)c|{\bf B}_c ,p_i,S_z\rangle \nonumber\\
&&=\bar{u}_{{\bf B}_n}(p_f,S_z')\left[ \gamma^+ f_1(k^2)-i\sigma^{+ \nu}\frac{k_{\nu}}{M_{{\bf B}_c}}f_2(k^2)+f_3(k^2)\frac{k^{+}}{M_{{\bf B}_c}}\right]u_{{\bf B}_c}(p_i,S_z)\nonumber\\
&&-\bar{u}_{{\bf B}_n}(p_f,S_z')\left[ \gamma^+ g_1(k^2)-i\sigma^{+ \nu}\frac{k_{\nu}}{M_{{\bf B}_c}}g_2(k^2)+g_3(k^2)\frac{k^{+}}{M_{{\bf B}_c}}\right]\gamma_5u_{{\bf B}_c}(p_i,S_z)
\end{eqnarray}
and choose the frame such that $p^+_{i(f)}$ is conserved ($k^+=0,k^2=-k_\perp^2$) to calculate the form factors to avoid  zero-mode contributions and
other $x^{+}$-ordered diagrams in the LF formalism~\cite{Schlumpf:1992ce,Zhang:1994ti}.
The Matrix elements of the vector and axial-vector currents at quark level correspond to three different lowest-order Feynman diagrams 
as shown in Fig.~1. 
\begin{figure}[b]
	\begin{minipage}[h]{0.3\linewidth}
		(a)
		\centering
		\includegraphics[width=2in]{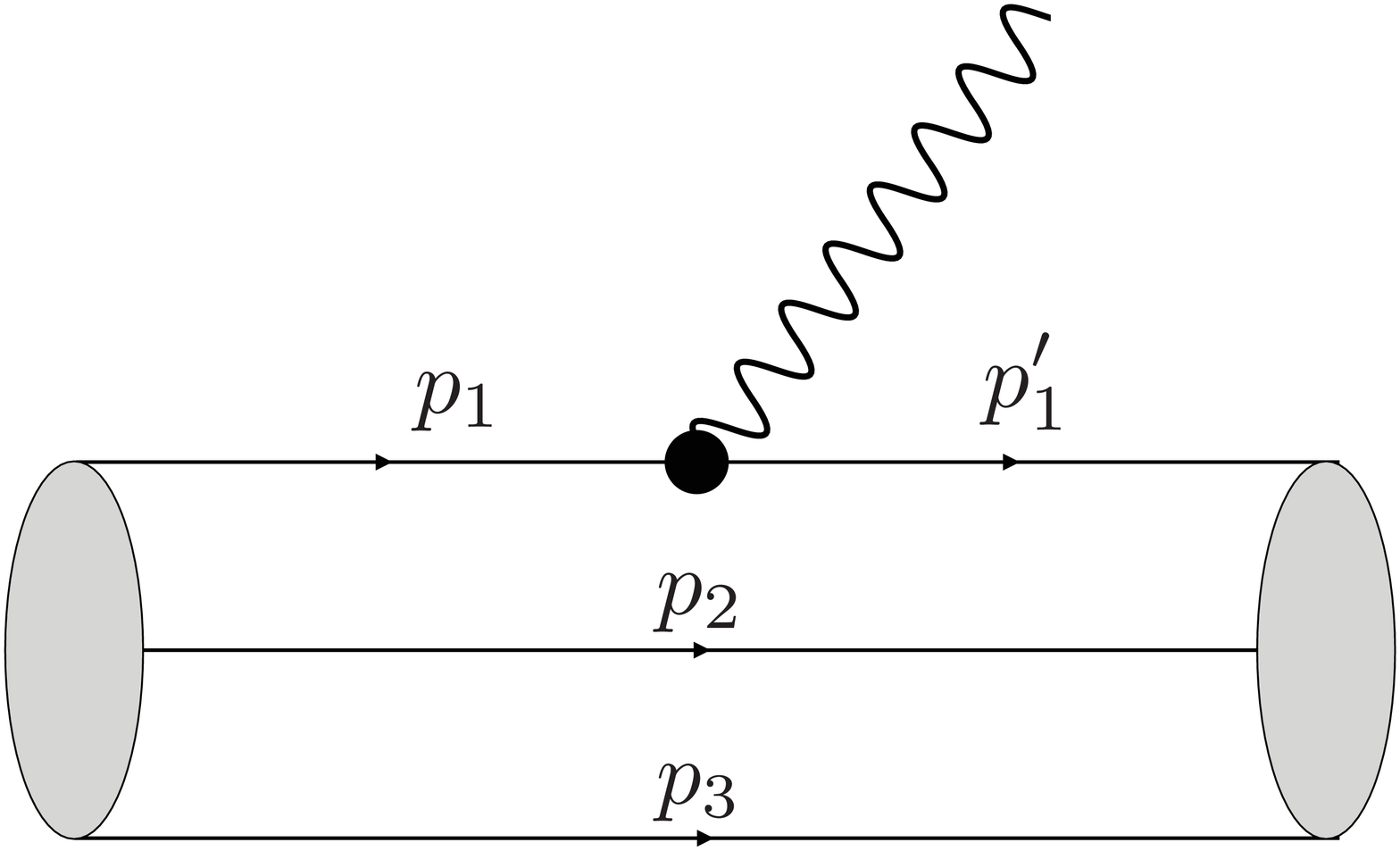}
	\end{minipage}
	\begin{minipage}[h]{0.3\linewidth}
		(b)
		\centering
		\includegraphics[width=2in]{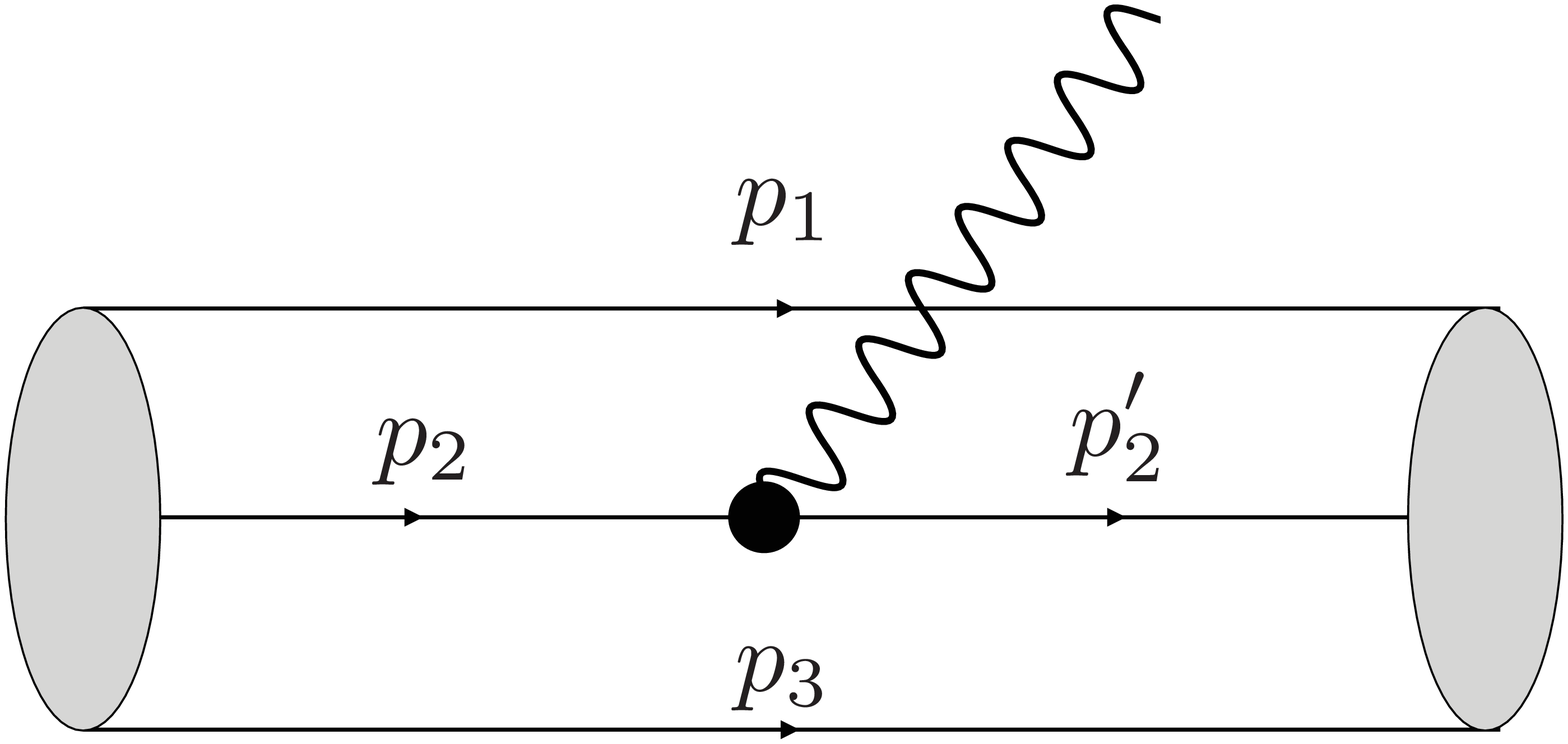}.
	\end{minipage}
	\begin{minipage}[h]{0.3\linewidth}
		(c)
		\centering
		\includegraphics[width=2in]{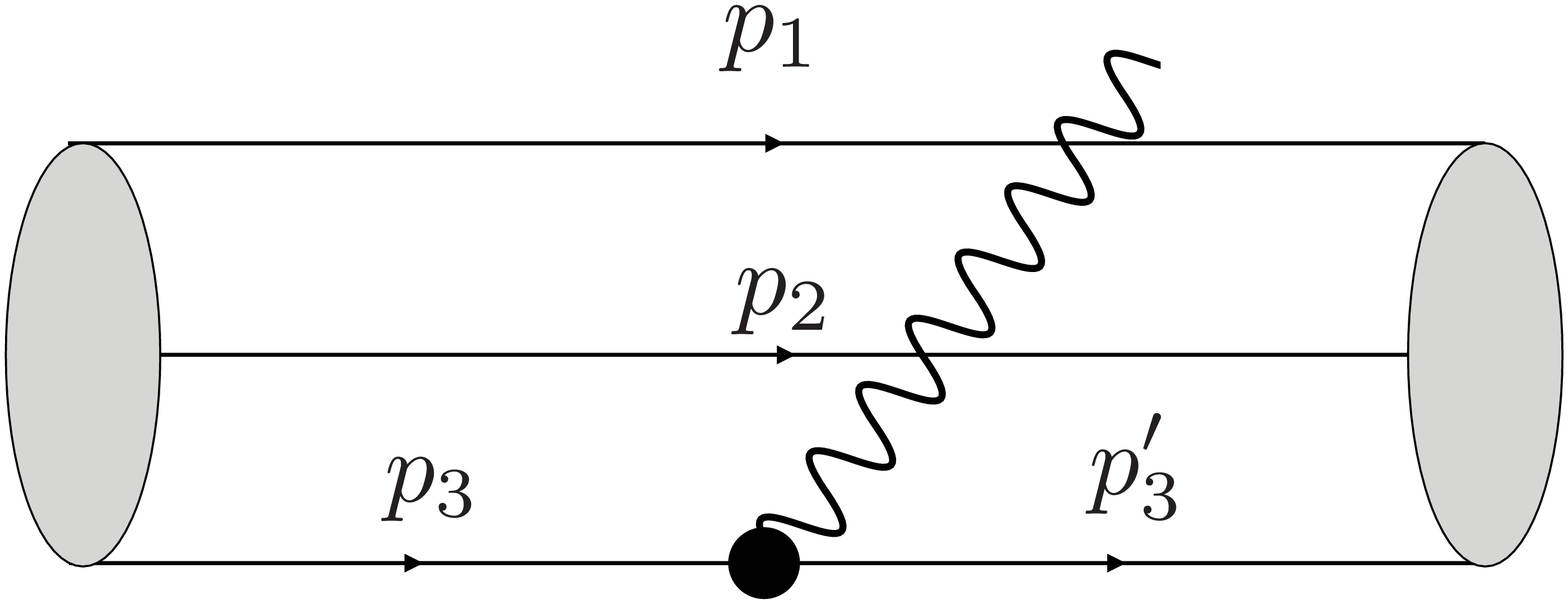}
	\end{minipage}
	\caption{Feynman diagrams for the baryonic weak transitions at the lowest order, where the sign of ``$\bullet$'' denotes the V-A  current vertex,
		where  (a) $p'_1-p_1=k$, (b) $p'_2-p_2=k$ and (c) $p'_3-p_3=k$.}
\end{figure}
Since the spin-flavor-momentum wave functions of baryons are totally symmetric under the permutation of constituents, 
we have that $(a)+(b)+(c)=3(a)=3(b)=3(c)$~\cite{Schlumpf:1992ce}. 
We only present the calculation for  the diagram (c), 
which contains simpler and cleaner forms with our notation $(q_{\perp},Q_{\perp},\xi,\eta)$ as a demonstration.
We can extract the form factors from the matrix elements through the relations
\begin{eqnarray}
	f_1(k^2)&=&\frac{1}{2p^+_i}\langle {\bf B_n},p_f,\uparrow|\bar{q}\gamma^{+}c|{\bf B_c},p_i,\uparrow\rangle\,, \nonumber \\
	f_2(k^2)&=&\frac{1}{2p^+_i}\frac{M_{{\bf B_c}}}{k_{\perp}}\langle {\bf B_n},p_f,\uparrow|\bar{q}\gamma^{+}c|{\bf B_c},p_i,\downarrow\rangle\,, \nonumber \\
	g_1(k^2)&=&\frac{1}{2p^+_i}\langle {\bf B_n},p_f,\uparrow|\bar{q}\gamma^{+}\gamma_5c|{\bf B_c},p_i,\uparrow\rangle\,, \nonumber \\
	g_2(k^2)&=&\frac{1}{2p^+_i}\frac{M_{{\bf B_c}}}{k_{\perp}}\langle {\bf B_n},p_f,\uparrow|\bar{q}\gamma^{+}\gamma_5|{\bf B_c},p_i,\downarrow\rangle  \,.\label{fm}
\end{eqnarray}
Note that $f_3$ and $g_3$ are not available  when $k^+=0$, but they are negligible because of the suppressions of  $k^2/M_{\bf B_c}^2$.
In fact, $f_3$ and $g_3$  are only associated with $H_{ \frac{1}{2}t}^{V(A)}$ which do not contribute to the semileptonic decays in 
the massless lepton limit~\cite{Geng:2019bfz}.
As a result,  we can safely set both $f_3$ and $g_3$  to be 0 in this study.
With the help of the  momentum distribution functions and  Melosh transformation matrix in Eq.~(\ref{baryon}),  
the transition matrix elements  can be written as
\begin{eqnarray}
&&\langle {\bf B_n},p_f,S'_z|\bar{q}\gamma^{+}(\gamma^5)c|{\bf B_c},p_i,S_z\rangle \nonumber \\
&&=\frac{1}{2^2(2\pi)^6}\int d\xi d\eta d^2q_\perp d^2Q_\perp\phi(q'_{\perp},\xi,Q'_\perp,\eta)\phi_{3}(q_{\perp},\xi,Q_\perp,\eta)F^{lmn} F_{ijk}\delta_{l}^{i}\delta_{m}^{j}\nonumber\\
&&\times \sum_{s_1,s_2,s_3}\sum_{s'_1,s'_2,s'_3}\langle S',S'_z|s'_1,s'_2,s'_3\rangle\langle s_1,s_2,s_3|S,S_z\rangle
\langle s'_1|R'_1R^{\dagger}_1|s_1\rangle\langle s'_2|R'_2R^{\dagger}_2|s_2\rangle \nonumber\\
&&\times 2P^+(3\delta_{n}^{q}\delta^{k}_{c})\langle s'_3|R'_3\sum_{\lambda'_3\lambda_3} \left(\delta^{\lambda_3}_{\lambda'_3}([\sigma_z]^{\lambda_3}_{\lambda'_3})|\lambda'_3\rangle \langle\lambda_3|\right) R_3^\dagger|s_3\rangle \label{vf}\,,
\end{eqnarray}
where the indices of $q$ and $c$ in the delta symbols correspond to the quark flavors in the $\bar{q}\gamma^{+}(\gamma^5)c$ current,
$q_{\perp}'=q_{\perp}$ and $Q_{\perp}'=Q_{\perp}+k_{\perp}$.
Using Eqs.~(\ref{fm}) and (\ref{vf}), we find that 
\begin{eqnarray}
\label{f1}
&&f_1(k^2)=\frac{3}{2^2(2\pi)^6}\int d\xi d\eta d^2q_\perp d^2Q_\perp\phi(q'_{\perp},\xi,Q'_\perp,\eta)\phi_3(q_{\perp},\xi,Q_\perp,\eta)(F^{lmn} F_{ijk}\delta_{n}^{q}\delta^{k}_{c}\delta_{l}^{i}\delta_{m}^{j})\nonumber\\
&&\times \sum_{s_1,s_2,s_3}\sum_{s'_1,s'_2,s'_3}\langle S',\uparrow|s'_1,s'_2,s'_3\rangle\langle s_1,s_2,s_3|S,\uparrow\rangle
\prod_{i=1,2,3}\langle s'_i|R'_iR^{\dagger}_i|s_i\rangle \,,
\end{eqnarray}
\begin{eqnarray}
\label{g1}
&&g_1(k^2)=\frac{3}{2^2(2\pi)^6}\int d\xi d\eta d^2q_\perp d^2Q_\perp\phi(q'_{\perp},\xi,Q'_\perp,\eta)\phi_3(q_{\perp},\xi,Q_\perp,\eta)(F^{lmn} F_{ijk}\delta_{n}^{q}\delta^{k}_{c}\delta_{l}^{i}\delta_{m}^{j})\nonumber\\
&&\times \sum_{s_1,s_2,s_3}\sum_{s'_1,s'_2,s'_3}\langle S',\uparrow|s'_1,s'_2,s'_3\rangle\langle s_1,s_2,s_3|S,\uparrow\rangle
\prod_{i=1,2}\langle s'_i|R'_iR^{\dagger}_i|s_i\rangle \langle s'_3|R'_3\sigma_zR^{\dagger}_3|s_3\rangle\,,
\end{eqnarray}
\begin{eqnarray}
\label{f2}
&&f_2(k^2)=\frac{3}{2^2(2\pi)^6}\frac{M_{{\bf B_c}}}{k_{\perp}}\int d\xi d\eta d^2q_\perp d^2Q_\perp\phi(q'_{\perp},\xi,Q'_\perp,\eta)\phi_3(q_{\perp},\xi,Q_\perp,\eta)(F^{lmn} F_{ijk}\delta_{n}^{q}\delta^{k}_{c}\delta_{l}^{i}\delta_{m}^{j})\nonumber\\
&&\times \sum_{s_1,s_2,s_3}\sum_{s'_1,s'_2,s'_3}\langle S',\uparrow|s'_1,s'_2,s'_3\rangle\langle s_1,s_2,s_3|S,\downarrow\rangle
\prod_{i=1,2,3}\langle s'_i|R'_iR^{\dagger}_i|s_i\rangle  \,,
\end{eqnarray}
\begin{eqnarray}
&&g_2(k^2)=\frac{3}{2^2(2\pi)^6}\frac{M_{{\bf B_c}}}{k_{\perp}}\int d\xi d\eta d^2q_\perp d^2Q_\perp\phi(q'_{\perp},\xi,Q'_\perp,\eta)\phi_3(q_{\perp},\xi,Q_\perp,\eta)(F^{lmn} F_{ijk}\delta_{n}^{q}\delta^{k}_{c}\delta_{l}^{i}\delta_{m}^{j})\nonumber\\
&&\times \sum_{s_1,s_2,s_3}\sum_{s'_1,s'_2,s'_3}\langle S',\uparrow|s'_1,s'_2,s'_3\rangle\langle s_1,s_2,s_3|S,\downarrow\rangle
\prod_{i=1,2}\langle s'_i|R'_iR^{\dagger}_i|s_i\rangle \langle s'_3|R'_3\sigma_zR^{\dagger}_3|s_3\rangle\,.
\label{g2}  
\end{eqnarray}

\section{Numerical Results}
To find out the decay branching ratios and averaged  asymmetries in the helicity formalism,
 we first calculate the  transition form factors with LFCQM.  
 By imposing the condition $k^+=0$, the form factors can be evaluated only in the space-like region ($k^2=-k^2_\perp$) 
 instead of the time-like one. Nonetheless, we still can extract the time-like information of the form factors via analytically continuations~\cite{Ke:2007tg,Cheng:2004cc,Jaus:1991cy,Jaus:1996np}. 

 We fit $f_{1(2)}(k^2)$ and $g_{1(2)}(k^2)$ with the analytic functions in the space-like region with the following form
 \begin{eqnarray}
 F(k^2)=\frac{F(0)}{1-q_1k^2+q_2k^4} \,.
 \end{eqnarray}
 We employ the numerical values of the constituent quark masses and shape parameters in Table.~I. 
 \begin{table}
\label{sh}
\caption{Values of the constituent quark masses ($m_i$) and shape parameters ($\beta_{q{\bf B_c}}$, $\beta_{{Q\bf B_c}},\beta_{{\bf B_n}}$) in units of GeV.} 
\begin{tabular}{cccccc}
\hline
$m_c$&$m_s$&$m_d$&$m_c$&$\beta_{q\Lambda_c}$&$\beta_{Q\Lambda_c}$\\
\hline
1.3&0.4&0.26&0.26&0.44&0.53\\
\hline
\hline
$\beta_{q\Xi_c}$&$\beta_{Q\Xi_c}$&$\beta_{N}$&$\beta_{\Lambda}$&$\beta_{\Sigma}$&$\beta_{\Xi}$\\
\hline
0.49&0.53&0.32&0.34&0.34&0.37\\
\hline
\end{tabular}
\end{table} 
 The values of the shape parameters can be determined approximately by the calculations in the mesonic cases~\cite{Chang:2018zjq,Ke:2019smy}.
   Because the strength of the quark-quark pairs is a half of the quark-anti-quark one~\cite{Ke:2019smy}, we will get the shape parameters of the quark pairs,
   which are approximately $\sqrt{2}$ smaller than those  in the mesonic cases.

    We adopt $\beta_{q\Lambda_c}\simeq2(\beta_{u\bar{d}}/\sqrt{2})$ and  $\beta_{q\Xi_c}\simeq2(\beta_{s\bar{u}(\bar{d})}/\sqrt{2})$,
   where the factor of 2 comes from the effects of the diquark clustering,  making the light quark pairs to be more compact in ${\bf B_c}$ baryons.  
   For the octet baryons of ${\bf B_n}$, we assume that the $SU(3)_f$ flavor symmetry is hold, and therefore, the shape parameters is flavor symmetric for each constituent,
   $i.e.$ $\beta_{Q}=\beta_q$. As a result, we approximate the shape parameters of the octet baryons equal to the mesonic ones by effectively treating any pair of two constituents as a heavier anti-quark.
  By using  Eqs.~(\ref{f1})-(\ref{g2}), we numerically compute 32 points for all form factors from $k^2=0$ to $k^2=-9.7\text{ GeV}^2$ and fit them with the MATLAB curve fitting toolbox. We present our fitting results of  the form factors in Tables~\ref{nLf},~\ref{nXipf} and~\ref{nXi0f}, and our predictions of 
  the branching ratios and asymmetry parameters in Table~\ref{result}.  The comparisons with different theoretical models  are presented
   in Tables~\ref{comLc}, \ref{comXip} and ~\ref{comXi0}, respectively.
\begin{table}
	\caption{Fitting results of the $\Lambda_c^+\to {\bf B_n}$ form factors in LFCQM}
	{
\begin{tabular}{ccccc}
	\hline
	\hline
	\multicolumn{5}{c}{$\Lambda_c^+ \to \Lambda$}\\
	\hline
	&$f_1$&$f_2$&$g_1$&$g_2$\\
	$F(0)$&$0.67\pm0.01$&$0.76\pm0.02$&$0.59\pm0.01$&$(3.8\pm1.2)\times10^{-3}$\\
	$q_1$&$1.48\pm0.31$&$1.44\pm0.30$&$1.22\pm0.28$&$4.99\pm17.6$\\
	$q_2$&$2.29\pm0.49$&$2.23\pm0.48$&$1.82\pm0.39$&$24.8\pm95.8$\\
		\hline\hline
	\multicolumn{5}{c}{$\Lambda_c^+ \to n$}\\
	\hline
	&$f_1$&$f_2$&$g_1$&$g_2$\\
	$F(0)$&$0.83\pm0.01$&$1.05\pm0.02$&$0.71\pm0.01$&$0.27\pm0.01$\\
	$q_1$&$1.25\pm0.36$&$1.20\pm0.30$&$0.94\pm0.28$&$1.37\pm0.40$\\
	$q_2$&$1.85\pm0.68$&$1.78\pm0.48$&$1.36\pm0.39$&$2.08\pm0.83$\\
	\hline
	\hline
\end{tabular}
\label{nLf}}
\end{table}
\begin{table}
	{\caption{Fitting results of the $\Xi_c^+\to {\bf B_n}$ form factors in LFCQM, }
	\begin{tabular}{ccccc}
		\hline
		\hline
		\multicolumn{5}{c}{$\Xi_c^+ \to \Xi^0$}\\
		\hline
		&$f_1$&$f_2$&$g_1$&$g_2$\\
		$F(0)$&$0.77\pm0.02$&$0.96\pm0.02$&$0.69\pm0.01$&$(6.8\pm0.3)\times10^{-3}$\\
		$q_1$&$1.50\pm0.31$&$1.45\pm0.31$&$1.25\pm0.28$&$2.00\pm0.76$\\
		$q_2$&$2.30\pm0.49$&$2.26\pm0.49$&$1.85\pm0.39$&$3.00\pm1.41$\\
		\hline\hline
		\multicolumn{5}{c}{$\Xi_c^+ \to \Sigma^0$}\\
		\hline
		&$f_1$&$f_2$&$g_1$&$g_2$\\
		$F(0)$&$0.52\pm0.01$&$0.70\pm0.02$&$0.45\pm0.01$&$0.08\pm0.01$\\
		$q_1$&$1.49\pm0.32$&$1.43\pm0.33$&$1.18\pm0.28$&$1.88\pm0.39$\\
		$q_2$&$2.35\pm0.51$&$2.38\pm0.53$&$1.79\pm0.38$&$2.88\pm0.67$\\
		\hline
		\hline
		\multicolumn{5}{c}{$\Xi_c^+ \to \Lambda$}\\
		\hline
		&$f_1$&$f_2$&$g_1$&$g_2$\\
		$F(0)$&$0.28\pm0.01$&$0.38\pm0.01$&$0.25\pm0.01$&$0.04\pm0.01$\\
	$q_1$&$1.50\pm0.31$&$1.35\pm0.51$&$1.18\pm0.28$&$1.71\pm0.38$\\
	$q_2$&$2.32\pm0.50$&$2.30\pm0.81$&$1.77\pm0.38$&$2.78\pm0.66$\\
		\hline
		\hline
	\end{tabular}
	\label{nXipf}}
\end{table}
\begin{table}
{	\caption{Fitting results of the $\Xi_c^0\to {\bf B_n}$ form factors in LFCQM, }
	\begin{tabular}{ccccc}
		\hline
		\hline
		\multicolumn{5}{c}{$\Xi_c^0 \to \Xi^-$}\\
		\hline
		&$f_1$&$f_2$&$g_1$&$g_2$\\
		$F(0)$&$0.74\pm0.02$&$0.96\pm0.02$&$0.69\pm0.01$&$(6.8\pm0.3)\times10^{-3}$\\
		$q_1$&$1.50\pm0.31$&$1.47\pm0.31$&$1.21\pm0.28$&$2.00\pm0.76$\\
		$q_2$&$2.30\pm0.50$&$2.25\pm0.48$&$1.98\pm0.39$&$3.00\pm1.41$\\
		\hline\hline
		\multicolumn{5}{c}{$\Xi_c^0 \to \Sigma^-$}\\
		\hline
		&$f_1$&$f_2$&$g_1$&$g_2$\\
		$F(0)$&$0.73\pm0.01$&$0.99\pm0.02$&$0.63\pm0.01$&$0.11\pm0.01$\\
		$q_1$&$1.49\pm0.32$&$1.43\pm0.33$&$1.18\pm0.28$&$1.88\pm0.39$\\
		$q_2$&$2.35\pm0.51$&$2.38\pm0.53$&$1.79\pm0.38$&$2.88\pm0.70$\\
		\hline
		\hline
	\end{tabular}
	\label{nXi0f}}
\end{table}

\begin{table}[h]
	\caption{Predictions of the decay branching ratios and asymmetry parameters}
	\begin{tabular}{c|cc|cc}
	\hline\hline
		&\multicolumn{2}{c|}{$\ell^+=e^+$}&\multicolumn{2}{c}{$\ell^+=\mu^+$}\\
		\hline
		&${\cal B}(\%)$&$\alpha$&${\cal B}(\%)$&$\alpha$\\
		$\Lambda_c^+\to \Lambda \ell^+ \nu_\ell$&$3.55\pm1.04$&$-0.97\pm0.03$&$3.40\pm1.02$&$-0.98\pm.0.02$\\
		$\Lambda_c^+\to n \ell^+ \nu_\ell$&$0.36\pm0.15$&$-0.96\pm0.04$&$0.34\pm0.15$&$-0.96\pm0.04$\\
		\hline
		$\Xi_c^+\to \Xi^0\ell^+\nu_\ell$&$11.3\pm3.35$&$-0.97\pm0.03$&$10.8\pm3.3$&$-0.97\pm0.03$\\
		$\Xi_c^+\to \Sigma^0\ell^+\nu_\ell$&$0.33\pm0.09$&$-0.98\pm0.01$&$0.31\pm0.09$&$-0.98\pm0.02$\\
		$\Xi_c^+\to \Lambda\ell^+\nu_\ell$&$0.12\pm0.04$&$-0.98\pm0.02$&$0.11\pm0.05$&$-0.98\pm0.02$\\
		\hline
		$\Xi_c^0\to \Xi^-\ell^+\nu_\ell$&$3.49\pm0.95$&$-0.98\pm0.02$&$3.34\pm0.94$&$-0.98\pm0.02$\\
		$\Xi_c^0\to \Sigma^-\ell^+\nu_\ell$&$0.22\pm0.06$&$-0.98\pm0.02$&$0.21\pm0.06$&$-0.98\pm0.02$\\
		\hline\hline
	\end{tabular}
	\label{result}
\end{table}

The uncertainties of our results mainly arise from the extrapolations by the analytical continuation method. The magnitudes
 of average asymmetry parameters predicted by LFCQM  almost reach their extremums in all channels. As shown in 
  Tables~\ref{comLc}-\ref{comXi0}, we obtain that  ${\cal B}(\Lambda_c^+\to\Lambda e^+\nu_e)=(3.55\pm1.04)\%$, ${\cal B}(\Xi_c^+\to\Xi^0 e^+\nu_e)=(11.3\pm3.4)\%$ and ${\cal B}(\Xi_c^0\to\Xi^- e^+\nu_e)=(3.49\pm0.95)\%$, which are all consistent with the current data from 
   PDG~\cite{Zyla:2020zbs} and the BELLE experiments~\cite{Li:2018qak,Li:2019atu}. Our branching ratios are also consistent with the predictions of the relativistic quark model (RQM)~\cite{Faustov:2016yza,Faustov:2019ddj}, the covariant constituent quark model (CCQM)~\cite{Gutsche:2015rrt} and the $SU(3)_F$ approach \cite{Geng:2019bfz}. The results given by the LF formalism~\cite{Zhao:2018zcb} and heavy quark effective theory (HQET)~\cite{Cheng:1995fe} are  half of ours because they choose the spin-flavor wave function of ${\bf B_c}$ to be $c(q_1q_2-q_2q_1)\chi_{s_z}^{\rho_3}$ instead of the totally symmetric one.
The averaged asymmetry parameter predicted by LFCQM  in  $\Lambda_c^+\to \Lambda e^+\nu_e$  is $10\%$ lower than the data. 
Since we do not consider any parameters about spin interactions  in our phenomenological wave functions, the helicity-structure-related results,
such as  the decay asymmetries, are clearly  not precise enough to explain the experimental data. On the other hand,  the prediction from  RQM and the $SU(3)_f$ approach are almost the same as the data.   Since the authors for RQM in Refs.~\cite{Faustov:2016yza,Faustov:2019ddj}
have considered a comprehensive QCD-inspired potential including the chromomagnetic effect, their results are more close to the experimental values than ours.   Meanwhile, the $SU(3)_f$ approach is a model independent way to analyze ${\bf B_c}\to {\bf B_n}\ell^+ \nu_{\ell}$ decays, which automatically 
includes the information related to the  spin interaction when  the asymmetry parameters are used as the fitting inputs. 
 Clearly, our results of the asymmetry parameters 
 could be improved by considering the full QCD potential and its solutions.
\begin{table}
	\caption{ Our results of $\Lambda_c^+\to {\bf B_n} e^+ \nu_e$ decays in comparisons with the experimental data and those in various calculations in the literature.}
	{	
		\begin{tabular}{c|cc|cc}
			\hline\hline
			&\multicolumn{2}{c|}{$\Lambda_c^+\to \Lambda e^+\nu_e$}&\multicolumn{2}{c}{$\Lambda_c^+\to n e^+\nu_e$}\\
			\hline
			&${\cal B}(\%)$&$\alpha$&${\cal B}(\%)$&$\alpha$\\
			LFCQM&$3.55\pm1.04$&$-0.97\pm0.03$&$0.36\pm0.15$&$-0.96\pm0.04$\\
			Data~\cite{Zyla:2020zbs}&$3.6\pm0.4$& $-0.86\pm0.04$&-&-\\
			$SU(3)$~\cite{Geng:2019bfz}&$3.2\pm0.3$&$-0.86\pm0.04$&$0.51\pm0.04$&$-0.89\pm0.04$\\
			RQM~\cite{Faustov:2016yza,Faustov:2019ddj}&3.25&-0.86&0.268&-0.91\\
			HQET~\cite{Cheng:1995fe}&1.42&-&-&-\\
			LF~\cite{Zhao:2018zcb}&1.63&-&0.201&-\\
			MBM~(NRQM)~\cite{PerezMarcial:1989yh}
			&2.6~(3.2)&-&0.20~(0.30)&-\\
			LQCD~\cite{Meinel:2016dqj,Meinel:2017ggx}&$3.80\pm0.22$&-&$0.41\pm0.03$&-\\
			CCQM~\cite{Gutsche:2015rrt}&2.78&-0.87&0.20&-\\
			LCSR~\cite{Liu:2009sn,Azizi:2009wn}&$3.0\pm0.3$&-&$8.69\pm2.89$&-\\
			\hline\hline	
		\end{tabular}
	\label{comLc}
	}
\end{table}

	\begin{table}
		{
			\centering
			\caption{ Our results of $\Xi_c^+\to {\bf B_n} e^+ \nu_e$ decays in comparisons with the experimental data and those in various calculations in the literature.}
			\begin{tabular}{c|cc|cc}
				\hline\hline
				&\multicolumn{2}{c|}{\multirow{2}{*}{$\Xi_c^+\to \Xi^0 e^+\nu_e$}}&\multicolumn{2}{c}{$\Xi_c^+\to \Sigma^0 e^+\nu_e$} \\
				&&&\multicolumn{2}{c}{$\Xi_c^+\to \Lambda e^+\nu_e$}\\
				\hline
				&${\cal B}(\%)$&$\alpha$&${\cal B}(\%)$&$\alpha$\\
				\multirow{2}{*}{LFCQM}&\multirow{2}{*}{$11.3\pm3.35$}&\multirow{2}{*}{$-0.97\pm0.03$}&$0.33\pm0.09$&$-0.98\pm0.01$\\
				&&&$0.12\pm0.04$&$-0.98\pm0.02$\\
				\hline
				\multirow{2}{*}{Data~\cite{Zyla:2020zbs,Li:2019atu}}&\multirow{2}{*}{$6.6^{+3.7}_{-3.5}$}& \multirow{2}{*}{-}&-&-\\
				&&&-&-\\
				\hline
				\multirow{2}{*}{$SU(3)$~\cite{Geng:2019bfz}}&\multirow{2}{*}{$10.7\pm0.9$}&\multirow{2}{*}{$-0.83\pm0.04$}&$0.46\pm0.04$&$-0.85\pm0.04$\\
				&&&$0.22\pm0.02$&$-0.86\pm0.04$\\
				\hline
				\multirow{2}{*}{RQM~\cite{Faustov:2019ddj}}&\multirow{2}{*}{9.40}&\multirow{2}{*}{-0.80}&-&-\\
				&&&0.13&-0.84\\
				\hline
				\multirow{2}{*}{LF~\cite{Zhao:2018zcb}}&\multirow{2}{*}{5.39}&\multirow{2}{*}{-}&0.19&-\\
				&&&0.08&-\\
				\hline
				\multirow{2}{*}{MBM~(NRQM)~\cite{PerezMarcial:1989yh}}
				&\multirow{2}{*}{11.1~(13.3)}&\multirow{2}{*}{-}&0.28~(0.41)&-\\
				&&&0.09~(0.14)&-\\
				\hline\hline	
			\end{tabular}
			\label{comXip}}
	\end{table}

\begin{table}
	\caption{ Our results of $\Xi_c^0\to{\bf B_n}e^+\nu_e$ decays in comparisons with the experimental data and those in various calculations in the literature.}
	\label{comXi0}
	{	
		\begin{tabular}{c|cc|cc}
			\hline\hline
			&\multicolumn{2}{c|}{$\Xi_c^0\to \Xi^- e^+\nu_e$}&\multicolumn{2}{c}{$\Xi_c^0\to \Sigma^- e^+\nu_e$}\\
			\hline
			&${\cal B}(\%)$&$\alpha$&${\cal B}(\%)$&$\alpha$\\
			LFCQM&$3.49\pm0.95$&$-0.98\pm0.03$&$0.22\pm0.06$&$-0.98\pm0.02$\\
			Data~\cite{Zyla:2020zbs,Li:2018qak}&$1.8\pm1.2$&-&-&-\\
			$SU(3)$~\cite{Geng:2019bfz}&$3.7\pm0.3$&$-0.83\pm0.04$&$0.33\pm0.03$&$-0.85\pm0.04$\\
			RQM~\cite{Faustov:2019ddj}&2.38&-0.80&-&--\\
			LF~\cite{Zhao:2018zcb}&1.35&-&0.10&-\\
			MBM~(NRQM)~\cite{PerezMarcial:1989yh}
			&3.55~(4.47)&-&0.19~(0.26)&-\\
			\hline\hline	
		\end{tabular}
	}
\end{table}

\section{Conclusions}
We have systematically studied the semi-leptonic decays of ${\bf B_c} \to {\bf B_n} \ell^+ \nu_{\ell}$ in  LFCQM.
By requiring the constituents in the baryonic states to obey the Fermi statistics, we are able to determine the overall spin-flavor-momentum structures of 
the baryons.
We assume that the momentum distribution of ${\bf B_n}$  is symmetric in flavor indices and any pair of two quarks can be effectively treated as a heavier anti-quark. 
We have found that 
${\cal B}(\Lambda_c^+\to\Lambda e^+\nu_e)=(3.55\pm1.04)\%$, ${\cal B}(\Xi_c^+\to\Xi^0 e^+\nu_e)=(11.3\pm3.35)\%$ and ${\cal B}(\Xi_c^0\to\Xi^- e^+\nu_e)=(3.49\pm0.95)\%$  in LFCQM, 
 which are consistent with the experimental data of  $(3.6\pm0.4)\times 10^{-2}$~\cite{Zyla:2020zbs}, $(6.6^{+3.7}_{-3.5})\times 10^{-2}$~\cite{Zyla:2020zbs,Li:2019atu} and $(1.8\pm1.2)\times 10^{-2}$~\cite{Zyla:2020zbs,Li:2018qak} as well as
 the values predicted by $SU(3)_F$~\cite{Geng:2019bfz}, LQCD~\cite{Meinel:2016dqj,Meinel:2017ggx},  
 RQM~\cite{Faustov:2016yza}, and CCQM~\cite{Gutsche:2015rrt}, but twice larger than those in HQET~\cite{Cheng:1995fe}
  and LF~\cite{Zhao:2018zcb}.
We have also obtained that 
${\alpha}( \Lambda_c^+ \to \Lambda e^+ \nu_{e})=(-0.97\pm0.03)$ in LFCQM,
which is $10\%$ lower than the experimental data of  $-0.86\pm0.04$~\cite{Zyla:2020zbs}.
The reason of  this deviation may arise from the QCD spin-spin interacting effects, which  are not included in our phenomenological wave functions.
 Our results of the averaged decay asymmetries could be improved if we consider the wave function solved from the full QCD potential. 
 It is clear that our predicted values for the decay branching ratios and asymmetries in $ \Xi_c^{+(0)} \to {\bf B_n} e^+ \nu_{e}$ 
could be tested in the ongoing experiments at  LHCb and  BELLEII.

\section*{ACKNOWLEDGMENTS}
We would like to thank Prof. Chengping Shen for discussions. This work was supported in part by National Center for Theoretical Sciences and 
MoST (MoST-107-2119-M-007-013-MY3).


\begin{thebibliography}{99}
	\bibitem{Li:2018qak} 
	Y.~B.~Li {\it et al.} [Belle Collaboration],
	Phys.\ Rev.\ Lett.\  {\bf 122},  082001 (2019).
	\bibitem{Li:2019atu} 
	Y.~B.~Li {\it et al.} [Belle Collaboration],
	Phys.\ Rev.\ D {\bf 100},  031101 (2019).
\bibitem{Aaij:2019lwg} 
R.~Aaij {\it et al.} [LHCb Collaboration],
Phys.\ Rev.\ D {\bf 100},  032001 (2019)
\bibitem{Zyla:2020zbs}
P.~A.~Zyla \textit{et al.} [Particle Data Group],
PTEP \textbf{2020},  083C01 (2020).


\bibitem{Savage:1989qr} 
M.~J.~Savage and R.~P.~Springer,
Phys.\ Rev.\ D {\bf 42}, 1527 (1990).

\bibitem{Savage:1991wu} 
M.~J.~Savage,
Phys.\ Lett.\ B {\bf 257}, 414 (1991).


\bibitem{Lu:2016ogy} 
C.~D.~Lu, W.~Wang and F.~S.~Yu,
Phys.\ Rev.\ D {\bf 93}, 056008 (2016).



\bibitem{Geng:2017mxn} 
C.~Q.~Geng, Y.~K.~Hsiao, C.~W.~Liu and T.~H.~Tsai,
JHEP {\bf 1711}, 147 (2017).

\bibitem{Wang:2017azm} 
W.~Wang, Z.~P.~Xing and J.~Xu,
Eur.\ Phys.\ J.\ C {\bf 77}, 800 (2017).


\bibitem{zero}
C.~Q.~Geng, Y.~K.~Hsiao, Y.~H.~Lin and L.~L.~Liu,
Phys.\ Lett.\ B {\bf 776}, 265 (2018).


\bibitem{Geng:2018plk} 
C.~Q.~Geng, Y.~K.~Hsiao, C.~W.~Liu and T.~H.~Tsai,
Phys.\ Rev.\ D {\bf 97},  073006 (2018);
Eur.\ Phys.\ J.\ C {\bf 78},  593 (2018).


\bibitem{Wang:2017gxe} 
D.~Wang, P.~F.~Guo, W.~H.~Long and F.~S.~Yu,
JHEP {\bf 1803}, 066 (2018).




\bibitem{He:2018php} 
X.~G.~He and W.~Wang,
Chin.\ Phys.\ C {\bf 42},  103108 (2018).


\bibitem{Geng:2018upx} 
C.~Q.~Geng, Y.~K.~Hsiao, C.~W.~Liu and T.~H.~Tsai,
Phys.\ Rev.\ D {\bf 99},  073003 (2019).

\bibitem{Geng:2018rse} 
C.~Q.~Geng, C.~W.~Liu and T.~H.~Tsai,
Phys.\ Lett.\ B {\bf 790}, 225 (2019);
Phys.\ Lett.\ B {\bf 794}, 19 (2019).

\bibitem{Geng:2019awr} 
C.~Q.~Geng, C.~W.~Liu, T.~H.~Tsai and Y.~Yu,
Phys.\ Rev.\ D {\bf 99},  114022 (2019).


\bibitem{Cen:2019ims} 
J.~Y.~Cen, C.~Q.~Geng, C.~W.~Liu and T.~H.~Tsai,
Eur.\ Phys.\ J.\ C {\bf 79},  946 (2019).

\bibitem{Hsiao:2019yur} 
Y.~K.~Hsiao, Y.~Yao and H.~J.~Zhao,
Phys.\ Lett.\ B {\bf 792}, 35 (2019).


\bibitem{Geng:2019bfz} 
C.~Q.~Geng, C.~W.~Liu, T.~H.~Tsai and S.~W.~Yeh,
Phys.\ Lett.\ B {\bf 792}, 214 (2019).

\bibitem{Grossman:2019xcj} 
Y.~Grossman and S.~Schacht,
JHEP {\bf 1907}, 020 (2019).

\bibitem{He:2018joe} 
X.~G.~He, Y.~J.~Shi and W.~Wang,
Eur.\ Phys.\ J.\ C {\bf 80},  359 (2020).


\bibitem{Roy:2019cky} 
S.~Roy, R.~Sinha and N.~G.~Deshpande,
Phys.\ Rev.\ D {\bf 101},  036018 (2020).

\bibitem{Geng:2020fng}
C.~Q.~Geng, C.~C.~Lih, C.~W.~Liu and T.~H.~Tsai,
Phys. Rev. D \textbf{101}, 094017 (2020).

\bibitem{Jia:2019zxi} 
C.~P.~Jia, D.~Wang and F.~S.~Yu,
Nucl.\ Phys.\ B {\bf 956}, 115048 (2020).

\bibitem{Cheng:1995fe} 
H.~Y.~Cheng and B.~Tseng,
Phys.\ Rev.\ D{\bf 53}, 1457 (1996).


\bibitem{Meinel:2016dqj} 
S.~Meinel,
Phys.\ Rev.\ Lett.\  {\bf 118}, 082001 (2017).
\bibitem{Meinel:2017ggx} 
S.~Meinel,
Phys.\ Rev.\ D{\bf 97}, 034511 (2018).
\bibitem{Faustov:2016yza} 
R.~N.~Faustov and V.~O.~Galkin,
Eur.\ Phys.\ J.\ C {\bf 76},  628 (2016).
\bibitem{Faustov:2019ddj}
R.~N.~Faustov and V.~O.~Galkin,
Eur. Phys. J. C \textbf{79},  695 (2019);
Particles \textbf{3},  208-222 (2020)

\bibitem{Gutsche:2015rrt} 
T.~Gutsche, M.~A.~Ivanov, J.~G.~Korner, V.~E.~Lyubovitskij and P.~Santorelli,
Phys.\ Rev.\ D {\bf 93}, 034008 (2016).
\bibitem{Liu:2009sn}
Y.~L.~Liu, M.~Q.~Huang and D.~W.~Wang,
Phys. Rev. D \textbf{80}, 074011 (2009)
\bibitem{Azizi:2009wn}
K.~Azizi, M.~Bayar, Y.~Sarac and H.~Sundu,
Phys. Rev. D \textbf{80}, 096007 (2009)
\bibitem{Zhao:2018zcb} 
Z.~X.~Zhao,
Chin.\ Phys.\ C {\bf 42},  093101 (2018).

\bibitem{PerezMarcial:1989yh} 
R.~Perez-Marcial, R.~Huerta, A.~Garcia and M.~Avila-Aoki,
Phys.\ Rev.\ D {\bf 40}, 2955 (1989)
Erratum: [Phys.\ Rev.\ D {\bf 44}, 2203 (1991)].


\bibitem{Zhang:1994ti} 
W.~M.~Zhang,
Chin.\ J.\ Phys.\  {\bf 32}, 717 (1994).
\bibitem{Schlumpf:1992ce} 
F.~Schlumpf,
hep-ph/9211255.


  
\bibitem{Geng:2000fs} 
  C.~Q.~Geng, C.~C.~Lih and W.~M.~Zhang,
  Phys.\ Rev.\ D {\bf 62}, 074017 (2000).
  
  

  
\bibitem{Geng:2001vy} 
  C.~Q.~Geng, C.~W.~Hwang and C.~C.~Liu,
  Phys.\ Rev.\ D {\bf 65}, 094037 (2002).
  
 



\bibitem{Cheng:2004cc} 
H.~Y.~Cheng, C.~K.~Chua and C.~W.~Hwang,
Phys.\ Rev.\ D {\bf 70}, 034007 (2004).

  
\bibitem{Geng:2016pyr} 
  C.~Q.~Geng, C.~C.~Lih and C.~Xia,
  Eur.\ Phys.\ J.\ C {\bf 76},  313 (2016).



\bibitem{Cheng:2017pcq} 
H.~Y.~Cheng and X.~W.~Kang,
Eur.\ Phys.\ J.\ C {\bf 77},  587 (2017)
Erratum: [Eur.\ Phys.\ J.\ C {\bf 77},  863 (2017)].


\bibitem{Zhao:2018mrg} 
Z.~X.~Zhao,
Eur.\ Phys.\ J.\ C {\bf 78},  756 (2018).
\bibitem{Xing:2018lre} 
Z.~P.~Xing and Z.~X.~Zhao,
Phys.\ Rev.\ D {\bf 98},  056002 (2018).
\bibitem{Chang:2019obq} 
Q.~Chang, L.~T.~Wang and X.~N.~Li,
JHEP {\bf 1912}, 102 (2019).


\bibitem{Chua:2018lfa} 
  C.~K.~Chua,
  Phys.\ Rev.\ D {\bf 99},  014023 (2019);
  Phys.\ Rev.\ D {\bf 100},  034025 (2019).

	\bibitem{Kadeer:2005aq}
	A.~Kadeer, J.~G.~Korner and U.~Moosbrugger,
	Eur. Phys. J. C \textbf{59}, 27-47 (2009)
\bibitem{Ke:2019smy} 
H.~W.~Ke, N.~Hao and X.~Q.~Li,
Eur.\ Phys.\ J.\ C {\bf 79},  540 (2019).
\bibitem{Ke:2007tg} 
H.~W.~Ke, X.~Q.~Li and Z.~T.~Wei,
Phys.\ Rev.\ D {\bf 77}, 014020 (2008).
\bibitem{Ke:2012wa} 
H.~W.~Ke, X.~H.~Yuan, X.~Q.~Li, Z.~T.~Wei and Y.~X.~Zhang,
Phys.\ Rev.\ D {\bf 86}, 114005 (2012).

\bibitem{Lorce:2011dv} 
C.~Lorce, B.~Pasquini and M.~Vanderhaeghen,
JHEP {\bf 1105}, 041 (2011).
\bibitem{Polyzou:2012ut} 
W.~N.~Polyzou, W.~Glockle and H.~Witala,
Few Body Syst.\  {\bf 54}, 1667 (2013).


\bibitem{Jaus:1991cy} 
W.~Jaus,
Phys.\ Rev.\ D {\bf 44}, 2851 (1991).
\bibitem{Jaus:1996np} 
W.~Jaus,
Phys.\ Rev.\ D {\bf 53}, 1349 (1996)
Erratum: [Phys.\ Rev.\ D {\bf 54}, 5904 (1996)].
\bibitem{Chang:2018zjq} 
Q.~Chang, X.~N.~Li, X.~Q.~Li, F.~Su and Y.~D.~Yang,
Phys.\ Rev.\ D {\bf 98},  114018 (2018).


\end{thebibliography}
\end{document}